\documentclass[preprint]{aastex}
\usepackage{graphicx}
\usepackage[utf8]{inputenc}   
\usepackage{plain}

\title{\textbf{VAMOS: a Pathfinder for the HAWC Gamma-Ray Observatory}}
\author{
\vspace*{2cm}
A. U. Abeysekara\altaffilmark{\ref{MSU}},
R. Alfaro\altaffilmark{\ref{IF-UNAM}},
C. Alvarez\altaffilmark{\ref{UNACH}},
J. D. {\'A}lvarez\altaffilmark{\ref{UMSNH}},
F. {\'A}ngeles\altaffilmark{\ref{IA-UNAM}},
R. Arceo\altaffilmark{\ref{UNACH}},
J. C. Arteaga-Vel{\'a}zquez\altaffilmark{\ref{UMSNH}},
A. Avila-Aroche\altaffilmark{\ref{IA-UNAM}},
H. A. Ayala Solares\altaffilmark{\ref{MTU}},
C. Badillo\altaffilmark{\ref{IF-UNAM}},
A. S. Barber\altaffilmark{\ref{University of Utah}},
B. M. Baughman\altaffilmark{\ref{UMD}},
N. Bautista-Elivar\altaffilmark{\ref{UPP}},
J. Becerra Gonzalez\altaffilmark{\ref{UMD},\ref{GSFC}},
E. Belmont\altaffilmark{\ref{IF-UNAM}},
E. Ben{\'i}tez\altaffilmark{\ref{IA-UNAM}},
S. Y. BenZvi\altaffilmark{\ref{UW-Madison}},
D. Berley\altaffilmark{\ref{UMD}},
A. Bernal\altaffilmark{\ref{IA-UNAM}},
M. Bonilla Rosales\altaffilmark{\ref{INAOE}},
J. Braun\altaffilmark{\ref{UMD}},
R. A. Caballero-Lopez\altaffilmark{\ref{IGeof-UNAM}},
K. S. Caballero-Mora\altaffilmark{\ref{CINVESTAV}},
I. Cabrera\altaffilmark{\ref{IF-UNAM}},
A. Carrami{\~n}ana\altaffilmark{\ref{INAOE}},
L. Casta{\~n}eda-Mart{\'i}nez\altaffilmark{\ref{IA-UNAM}},
M. Castillo\altaffilmark{\ref{FCFM-BUAP}},
U. Cotti\altaffilmark{\ref{UMSNH}},
J. Cotzomi\altaffilmark{\ref{FCFM-BUAP}},
E. de la Fuente\altaffilmark{\ref{UdG}},
C. De Le{\'o}n\altaffilmark{\ref{UMSNH}},
T. DeYoung\altaffilmark{\ref{PSU}},
A. Diaz-Azuara\altaffilmark{\ref{IA-UNAM}},
L. Diaz-Cruz\altaffilmark{\ref{FCFM-BUAP}},
R. Diaz Hernandez\altaffilmark{\ref{INAOE}},
J. C. D{\'\i}az-V{\'e}lez\altaffilmark{\ref{UW-Madison}},
B. L. Dingus\altaffilmark{\ref{LANL}},
D. Dultzin\altaffilmark{\ref{IA-UNAM}},
M. A. DuVernois\altaffilmark{\ref{UW-Madison}},
R. W. Ellsworth\altaffilmark{\ref{GMU},\ref{UMD}},
A. Fernandez\altaffilmark{\ref{FCFM-BUAP}},
D. W. Fiorino\altaffilmark{\ref{UW-Madison}},
N. Fraija\altaffilmark{\ref{IA-UNAM}},
A. Galindo\altaffilmark{\ref{INAOE}},
G. Garc{\'\i}a-Torales\altaffilmark{\ref{UdG}},
F. Garfias\altaffilmark{\ref{IA-UNAM}},
A. Gonz{\'a}lez\altaffilmark{\ref{IF-UNAM}},
L. X. Gonz{\'a}lez\altaffilmark{\ref{IGeof-UNAM}},
M. M. Gonz{\'a}lez\altaffilmark{\ref{IA-UNAM}},
J. A. Goodman\altaffilmark{\ref{UMD}},
V. Grabski\altaffilmark{\ref{IF-UNAM}},
M. Gussert\altaffilmark{\ref{CSU}},
C. Guzm{\'a}n-Cer{\'o}n\altaffilmark{\ref{IA-UNAM}},
Z. Hampel-Arias\altaffilmark{\ref{UW-Madison}},
J. P. Harding\altaffilmark{\ref{LANL}},
L. Hern{\'a}ndez-Cervantes\altaffilmark{\ref{IA-UNAM}},
C. M. Hui\altaffilmark{\ref{MTU}},
P. H{\"u}ntemeyer\altaffilmark{\ref{MTU}},
A. Imran\altaffilmark{\ref{UW-Madison}},
A. Iriarte\altaffilmark{\ref{IA-UNAM}},
P. Karn\altaffilmark{\ref{UC Irvine}},
D. Kieda\altaffilmark{\ref{University of Utah}},
G. J. Kunde\altaffilmark{\ref{LANL}},
R. Langarica\altaffilmark{\ref{IA-UNAM}},
A. Lara\altaffilmark{\ref{IGeof-UNAM}},
G. Lara\altaffilmark{\ref{IA-UNAM}},
R. J. Lauer\altaffilmark{\ref{UNM}},
W. H. Lee\altaffilmark{\ref{IA-UNAM}},
D. Lennarz\altaffilmark{\ref{GA Tech}},
H. Le{\'o}n Vargas\altaffilmark{\ref{IF-UNAM}},
E. C. Linares\altaffilmark{\ref{UMSNH}},
J. T. Linnemann\altaffilmark{\ref{MSU}},
M. Longo\altaffilmark{\ref{CSU}},
R. Luna-Garcia\altaffilmark{\ref{CIC-IPN}},
A. Marinelli*\altaffilmark{,\ref{IF-UNAM}},
L. A. Mart{\'\i}nez\altaffilmark{\ref{IA-UNAM}},
H. Mart{\'\i}nez\altaffilmark{\ref{CINVESTAV}},
O. Mart{\'\i}nez\altaffilmark{\ref{FCFM-BUAP}},
J. Mart{\'\i}nez-Castro\altaffilmark{\ref{CIC-IPN}},
M. Martos\altaffilmark{\ref{IA-UNAM}},
J. A. J. Matthews\altaffilmark{\ref{UNM}},
J. McEnery\altaffilmark{\ref{GSFC}},
E. Mendoza Torres\altaffilmark{\ref{INAOE}},
P. Miranda-Romagnoli\altaffilmark{\ref{UAEH}},
E. Moreno\altaffilmark{\ref{FCFM-BUAP}},
M. Mostaf{\'a}\altaffilmark{\ref{PSU}},
J. Nava\altaffilmark{\ref{INAOE}},
L. Nellen\altaffilmark{\ref{ICN-UNAM}},
M. Newbold\altaffilmark{\ref{University of Utah}},
R. Noriega-Papaqui\altaffilmark{\ref{UAEH}},
T. Oceguera-Becerra\altaffilmark{\ref{UdG}},
D. P. Page\altaffilmark{\ref{IA-UNAM}},
B. Patricelli\altaffilmark{\ref{IA-UNAM}},
R. Pelayo\altaffilmark{\ref{CIC-IPN}},
E. G. P{\'e}rez-P{\'e}rez\altaffilmark{\ref{UPP}},
J. Pretz\altaffilmark{\ref{PSU}},
I. Ram{\'i}rez\altaffilmark{\ref{IF-UNAM}},
A. Renter{\'i}a\altaffilmark{\ref{IF-UNAM}},
C. Rivi{\`e}re\altaffilmark{\ref{IA-UNAM}},
D. Rosa-Gonz{\'a}lez\altaffilmark{\ref{INAOE}},
F. Ruiz-Sala\altaffilmark{\ref{IA-UNAM}},
E. L. Ruiz-Velasco\altaffilmark{\ref{IA-UNAM}},
J. Ryan\altaffilmark{\ref{New Hampshire}},
J. R. Sacahui\altaffilmark{\ref{IA-UNAM}},
H. Salazar\altaffilmark{\ref{FCFM-BUAP}},
F. Salesa\altaffilmark{\ref{PSU}},
A. Sandoval\altaffilmark{\ref{IF-UNAM}},
E. Santos\altaffilmark{\ref{UNACH}},
M. Schneider\altaffilmark{\ref{UC Santa Cruz}},
S. Silich\altaffilmark{\ref{INAOE}},
G. Sinnis\altaffilmark{\ref{LANL}},
A. J. Smith\altaffilmark{\ref{UMD}},
K. Sparks Woodle\altaffilmark{\ref{PSU}},
R. W. Springer\altaffilmark{\ref{University of Utah}},
F. Suarez\altaffilmark{\ref{IF-UNAM}},
I. Taboada\altaffilmark{\ref{GA Tech}},
A. Tepe\altaffilmark{\ref{GA Tech}},
P. A. Toale\altaffilmark{\ref{UA}},
K. Tollefson\altaffilmark{\ref{MSU}},
I. Torres\altaffilmark{\ref{INAOE}},
S. Tinoco\altaffilmark{\ref{IA-UNAM}},
T. N. Ukwatta\altaffilmark{\ref{MSU}},
J. F. Vald{\'e}s Galicia\altaffilmark{\ref{IGeof-UNAM}},
P. Vanegas\altaffilmark{\ref{IF-UNAM}},
A. V{\'a}zquez\altaffilmark{\ref{IF-UNAM}},
L. Villase{\~n}or\altaffilmark{\ref{UMSNH}},
W. Wall\altaffilmark{\ref{INAOE}},
T. Weisgarber\altaffilmark{\ref{UW-Madison}},
S. Westerhoff\altaffilmark{\ref{UW-Madison}},
I. G. Wisher\altaffilmark{\ref{UW-Madison}},
J. Wood\altaffilmark{\ref{UMD}},
G. B. Yodh\altaffilmark{\ref{UC Irvine}},
P. W. Younk\altaffilmark{\ref{LANL}},
D. Zaborov**\altaffilmark{,\ref{PSU}},
A. Zepeda\altaffilmark{\ref{CINVESTAV}},
H. Zhou\altaffilmark{\ref{MTU}}}
\email{}
\email{}
\email{}
\email{}
\email{\hspace{-10.2 cm}\rule{6 cm}{0.01 cm}}
\email{\hspace{-11cm}\small\textrm{*antonio.marinelli@fisica.unam.mx}}
\email{\hspace{-12.6cm}\small\textrm{**zaborov@phys.psu.edu}}
\altaffiltext{a}{\label{MSU} Department of Physics and Astronomy, Michigan State University, East Lansing, MI, USA}
\altaffiltext{b}{\label{IF-UNAM} Instituto de F{\'\i}sica, Universidad Nacional Aut{\'o}noma de M{\'e}xico, Mexico D.F., Mexico}
\altaffiltext{c}{\label{UNACH} CEFyMAP, Universidad Aut{\'o}noma de Chiapas, Tuxtla Guti{\'e}rrez, Chiapas, Mexico}
\altaffiltext{d}{\label{UMSNH} Universidad Michoacana de San Nicol{\'a}s de Hidalgo, Morelia, Mexico}
\altaffiltext{e}{\label{IA-UNAM} Instituto de Astronom{\'\i}a, Universidad Nacional Aut{\'o}noma de M{\'e}xico, Mexico D.F., Mexico}
\altaffiltext{f}{\label{MTU} Department of Physics, Michigan Technological University, Houghton, MI, USA}
\altaffiltext{g}{\label{University of Utah} Department of Physics and Astronomy, University of Utah, Salt Lake City, UT, USA}
\altaffiltext{h}{\label{UMD} Department of Physics, University of Maryland, College Park, MD, USA}
\altaffiltext{i}{\label{UPP} Universidad Politecnica de Pachuca, Pachuca, Hgo, Mexico}
\altaffiltext{j}{\label{GSFC} Astrophysics Science Division, NASA Goddard Space Flight Center, Greenbelt, MD, USA}
\altaffiltext{k}{\label{UW-Madison} Department of Physics, University of Wisconsin-Madison, Madison, WI, USA}
\altaffiltext{l}{\label{INAOE} Instituto Nacional de Astrof{\'\i}sica, {\'O}ptica y Electr{\'o}nica, Tonantzintla, Puebla, Mexico}
\altaffiltext{m}{\label{IGeof-UNAM} Instituto de Geof{\'\i}sica, Universidad Nacional Aut{\'o}noma de M{\'e}xico, Mexico D.F., Mexico}
\altaffiltext{n}{\label{CINVESTAV} Physics Department, Centro de Investigacion y de Estudios Avanzados del IPN, Mexico D.F., Mexico}
\altaffiltext{o}{\label{FCFM-BUAP} Facultad de Ciencias F{\'\i}sico Matem{\'a}ticas, Benem{\'e}rita Universidad Aut{\'o}noma de Puebla, Puebla, Mexico}
\altaffiltext{p}{\label{UdG} Dept. de Fisica, Dept. de Electronica (CUCEI), IT.Phd (CUCEA), Phys-Mat. Phd (CUVALLES), Universidad \hspace*{0.55cm} de Guadalajara, Jalisco, Mexico}
\altaffiltext{q}{\label{PSU} Department of Physics, Pennsylvania State University, University Park, PA, USA}
\altaffiltext{r}{\label{LANL} Physics Division, Los Alamos National Laboratory, Los Alamos, NM, USA}
\altaffiltext{s}{\label{GMU} School of Physics, Astronomy, and Computational Sciences, George Mason University, Fairfax, VA, USA}
\altaffiltext{t}{\label{CSU} Physics Department, Ft Collins, CO 80523, USA}
\altaffiltext{u}{\label{UC Irvine} Department of Physics and Astronomy, University of California, Irvine, Irvine, CA, USA}
\altaffiltext{v}{\label{UNM} Dept of Physics and Astronomy, University of New Mexico, Albuquerque, NM, USA}
\altaffiltext{w}{\label{GA Tech} School of Physics and Center for Relativistic Astrophysics - Georgia Institute of Technology, Atlanta, GA,  USA \hspace*{0.55cm} 30332}
\altaffiltext{x}{\label{CIC-IPN} Centro de Investigacion en Computacion, Instituto Politecnico Nacional, Mexico D.F., Mexico }
\altaffiltext{y}{\label{UAEH} Universidad Aut{\'o}noma del Estado de Hidalgo, Pachuca, Hidalgo, Mexico}
\altaffiltext{z}{\label{ICN-UNAM} Instituto de Ciencias Nucleares, Universidad Nacional Aut{\'o}noma de M{\'e}xico, Mexico D.F., Mexico}
\altaffiltext{aa}{\label{New Hampshire} Space Science Center, University of New Hampshire, Durham, NH, USA }
\altaffiltext{ab}{\label{UC Santa Cruz} Santa Cruz Institute for Particle Physics, University of California, Santa Cruz, Santa Cruz, CA, USA}
\altaffiltext{ac}{\label{UA} Department of Physics \& Astronomy, University of Alabama, Tuscaloosa, AL, USA}

\pagebreak
\begin{document}

\pagebreak

\section*{Abstract}
VAMOS\footnote{VAMOS is an acronym for: Verification and Assessment Measuring of Observatory Subsystem} was a prototype detector built in 2011 at an altitude of 4100\,m a.s.l. in the state of 
Puebla, Mexico. The aim of VAMOS was to finalize the design, construction techniques and data acquisition system of the HAWC observatory. HAWC is an air-shower array currently under 
construction at the same site of VAMOS with the purpose to study the TeV sky. The VAMOS setup included six water Cherenkov detectors and two different data acquisition systems. It was in 
operation between October 2011 and May 2012 with an average live time of $30\%$. Besides the scientific verification purposes, the eight months of data were used to obtain the results presented in 
this paper:  the detector response to the Forbush decrease of March 2012, and the analysis of possible emission, at energies above 30\,GeV, for long gamma-ray bursts GRB111016B and 
GRB120328B.

\section*{Keywords}
\hspace{2.5cm}Detector prototype; Scientific verification; TeV cosmic rays
\section{Introduction}
The High Altitude Water Cherenkov (HAWC) observatory is an air-shower array composed of 300 water Cherenkov detectors (WCDs) currently under construction within the Pico de Orizaba National 
Park (Mexico) by a joint Mexican-United States collaboration, with planned completion in 2014. The main purpose of HAWC is to observe, with a large duty-cycle the TeV gamma-ray sky. 
For this purpose, the HAWC collaboration prototyped a new array architecture taking advantage of the expertise obtained with the Milagro 
\citep{milagro2006}  experiment. Milagro was a previous generation experiment that operated at a lower altitude of 2630\,m (respect to the 4100\,m of HAWC) close to Los Alamos, New Mexico. The 
pond design of the Milagro experiment has been replaced in the HAWC experiment by a modular design (array of WCDs). The WCD is designed to detect the Cherenkov light 
produced by secondary particles \citep{Stecker} using photomultiplier tubes (PMTs) immersed in water. While the overall layout of HAWC was redesigned to obtain an 
instrument fifteen times more sensitive than Milagro (for a point-like gamma-ray source with a Crab-like spectrum \citep{sensi}), the front-end electronics and the 8-inch PMTs have been inherited from 
the Milagro experiment. The higher elevation, larger area of the array, better optical isolation of the PMTs, and the introduction of a new 10-inch PMT in each WCD are the most important 
improvements of the HAWC array with respect to the Milagro experiment. To test the WCD design and  the main parts of the electronics, a small prototype called VAMOS was built. It was composed of 
six WCDs containing a total of 36 PMTs. The construction of VAMOS started in May 2010 and finished in June 2011, with the first data taken on June 17, 2011. VAMOS took data continuously from 
October 2011 through May 2012. \\ The first two sections of this paper are dedicated to the description of the main components of VAMOS, while in the following sections we report on analyses of the 
eight months of VAMOS data. In sections 4 and 5 we describe two analyses obtained with the scaler \citep{HAWCgrb} data acquisition (DAQ) system:  the effects of environmental variability and the 
effects of solar activity on the measured rate of events. In section 6 we report the validation of data and experiment design obtained by the comparison with Monte Carlo simulations. Finally, section 7 
is dedicated to the calculation of upper limits on the emission of GRB111016B  and GRB120328B considering the energy range from 30\,GeV up to few hundreds of GeV.

                                                                                                                                                                                                                                                                                                                                                                                                                                                                                                                                                                                                                                                                                                                                                                                                                                                \section{Water Cherenkov Detector}
\subsection{Site}
The VAMOS prototype was situated 170\,m Northwest from the center of HAWC, which is being constructed on the flank of the Sierra Negra volcano, central Mexico (N $18^{\circ}$59'41'', W 
$97^{\circ}$18'27''), at 4100\,m a.s.l. The Pico de Orizaba volcano, also known as Citlaltepetl, is the highest mountain in Mexico and towers over the site at 
5610\,m, 6\,km to the Northeast. The site is a plateau of soft volcanic soil and loose rocks which provides a suitable foundation for the construction of the WCDs. 
The temperature at the HAWC site is relatively mild; the all-year average temperature is $4.3^{\circ}$C \citep{Carrasco}, with sub-zero temperatures observed 
only 5\% of the time and mostly only early morning hours in the first part of the winter season. These conditions present no risk of WCD water freezing and thus 
ensure a homogeneous refractive index inside the WCDs during the whole year. The annual precipitation is around 1\,m and the relative humidity is about 50\% during the dry season
and 85\% during the wet season \citep{Carrasco}. The average wind speed is 4\,km/h \citep{Carrasco} at the top of Sierra Negra with occasional conditions of extreme weather: for example 
hurricane Dean generated winds of 140 km/h \citep{Carrasco} on August 22nd 2007 and hurricane Ernesto winds of 90\,km/h on August 8th 2012. The Sierra Negra is an extinct volcano while 
the Citlaltepetl is an active stratovolcano with the last major activity having occurred in 1545 CE. Geophysical studies \citep{DelaCruzReyna2002} give to the latter a probability of a minor 
explosive event as 0.013 per year ($\sim$13\% in 10 years of HAWC operations).  Earthquakes of magnitude 7 or larger have been felt in Ciudad Serdan, a city in the valley, 15\,km from the 
HAWC site, in 1937, 1973 and 1980.  Considering the reported environmental risks, the VAMOS design (as well as the HAWC design) incorporates features to withstand high winds and large 
seismicevents. The Large Millimeter Telescope (LMT) \citep{Yun} is a radio telescope situated at the summit of Sierra Negra that provided the access road, electricity and optical fiber that are 
now common infrastructure shared between several observatories on the mountain. In July 2009, a 1\,km long extension to the LMT road was constructed to access the VAMOS/HAWC site. 

\subsection{Tanks}
\begin{figure}[h!]
\centering
\begin{tabular}{cc}
\includegraphics[width=0.45\textwidth]{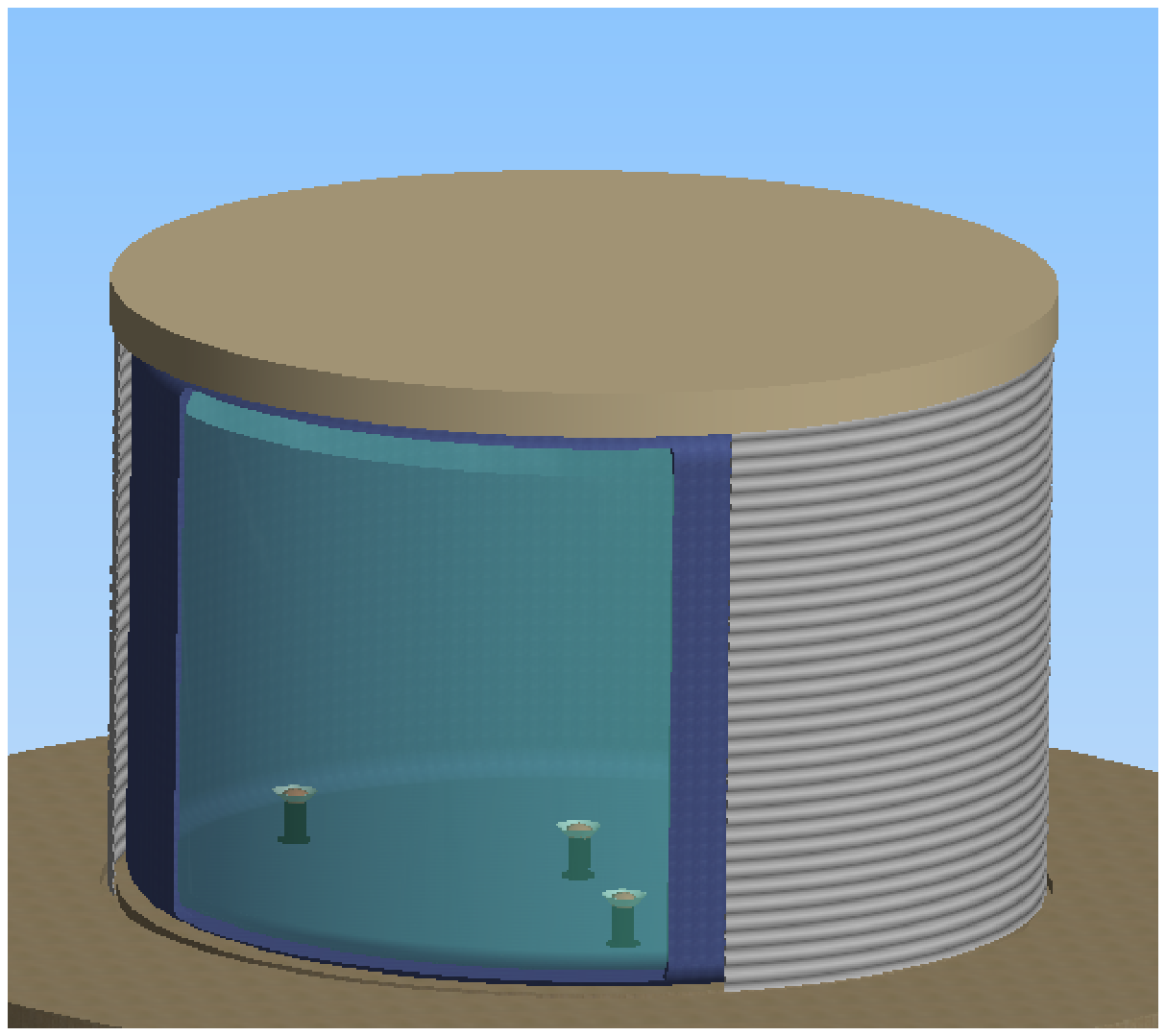}  &
\includegraphics[width=0.45\textwidth]{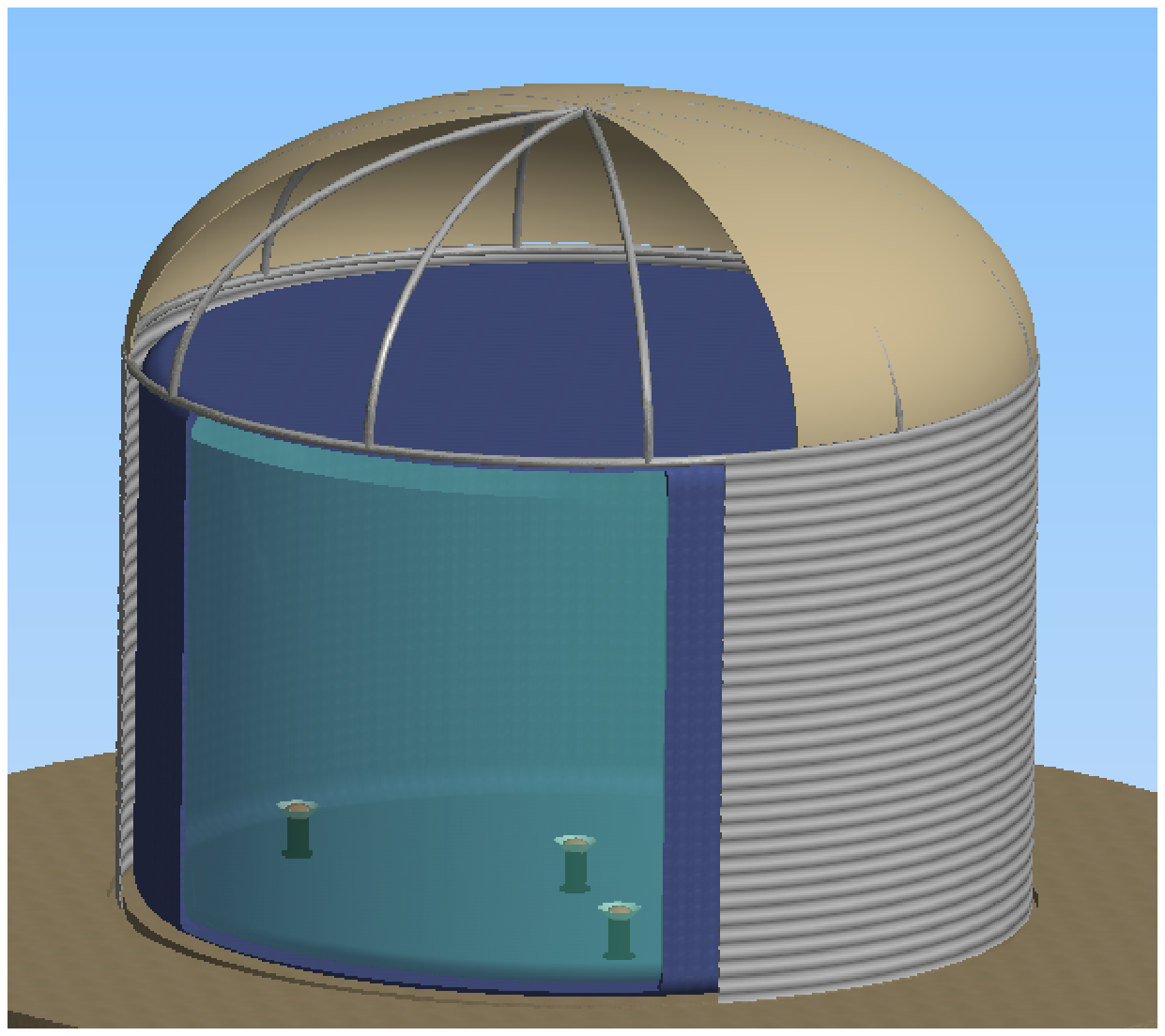} \\
\end{tabular}
\caption{On the left side a VAMOS WCD, with a height of 5\,m and a diameter of 7.3\,m while, on the right side, a HAWC WCD with the same  
dimensions and the introduction of a roof 0.6\,m high in the center of the WCD instead of a tarp. Both representations show the external steel tank (gray), the PVC 
bladder (dark blue), 4.2\,m of purified water (light green) and a sample of the PMTs anchored at the bottom of the WCD with their respective cone-shaped baffles.}\label{tank}
\end{figure}

The VAMOS WCDs were constructed using external steel tanks containing the light-tight  bladders which hold the purified water, the PMTs, and the high voltage cables (fig.~\ref{tank}). Before the 
construction of the VAMOS prototype, two designs were considered for the external structure of a WCD: a plastic molded water tank and a steel tank lined with a bladder. After several tests at the site 
and a study done with Monte Carlo simulation, the final design for the VAMOS prototype as well as for the HAWC array resulted in the choice of a the steel tank \citep{Sandoval} lined with a bladder. 
The  main advantages to use steel tanks were: the cost and the ease of transportation and assembly at the site. The steel tanks selected (fig.~\ref{tank}) have a diameter of 7.3\,m and a height of 5.0\,m. 
These dimensions were considered appropriate to discriminate the signal of gamma-ray induced air showers from the background of cosmic-ray induced air showers. In fact,                                                                                                                                                                                                                                                                                                                                                                                                                                                                                                                                                                                                                                                                                                                                                                                                                                       
smaller dimensions can compromise the ability to discriminate hadrons using large hits produced by muons passing through the WCDs, while larger tanks are less practical to construct and lead to a 
less granular detector which tends to worsen the angular resolution. The cylindrical tanks were constructed from galvanized steel that is commonly used for seed storage. The bottom 0.3\,m of the tank 
wall was buried in the ground to provide a natural anchor for stability and earthquake certification. The PMTs were 0.5\,m high including the encapsulated base; therefore the total amount of water 
above each PMT was 4.2\,m in depth or 10.5 radiation lengths. When filled, each tank contains about 188,000 liters of purified water. The depth of the water inside three of the 
VAMOS tanks was monitored with dedicated depth sensors.
\subsection{Bladders}
Bladders were custom made for the VAMOS tanks using 15\,mm thick black PVC of the type XR3 PW. The VAMOS bladders had anchor points at the top for fixing to the steel structure, 
seven PMT mounting points at the bottom, and a hatch at the top for access to deploy PMTs and to fill with water. The design of the bladders 
allowed easy access to the interior when the PMTs were installed. The top of the tank was covered with a tarp to protect the bladder from sun, rain, and wind. In HAWC 
tanks, the tarp is replaced by a vaulted roof with 0.6\,m of height to avoid the water or ice accumulation at the top of the WCD. The VAMOS 
bladders were tested for light leaks using a PMT from the Milagro experiment. A representation of VAMOS and HAWC WCD with bladders is shown in fig.~\ref{tank}.
The bladder material has also been tested for water contamination. This was a critical issue because there was no recirculation plan for the water of the tanks. The attenuation 
length of water when in contact with bladder material for the light at 325\,nm was observed to remain constant with time. The VAMOS experience was used to optimize the design of the bladders for 
the HAWC array. The bladder material was certified food safe, therefore no additional filtering will be needed before returning the water to the environment when the HAWC experiment will be over.
\subsection{Water delivery system} 
VAMOS was an important test  for obtaining and managing water near the site, allowing the optimization of water plans required by HAWC (60 million liters of filtered 
water are needed for the complete array). Elements of the Milagro water purification system have been used for VAMOS with the same filtration setup and a capacity of 370\,liters per minute. The 
water, obtained from a well 20\,km away from the VAMOS site, was filtered with a series of progressive stages using polypropylene filter cartridges. 
In particular, the water has been processed with two preliminary filtration stages of 10 and 5 microns before the finest one of 1 micron. Afterwards the water was sterilized
using a UV light source. The initial attenuation length obtained with this filtration system was calculated to be around 8\,m at 325\,nm.

\section{Instrumentation, electronics and data acquisition system}
\begin{figure}[!htb]
 \centering
 \includegraphics[width=0.6\textwidth]{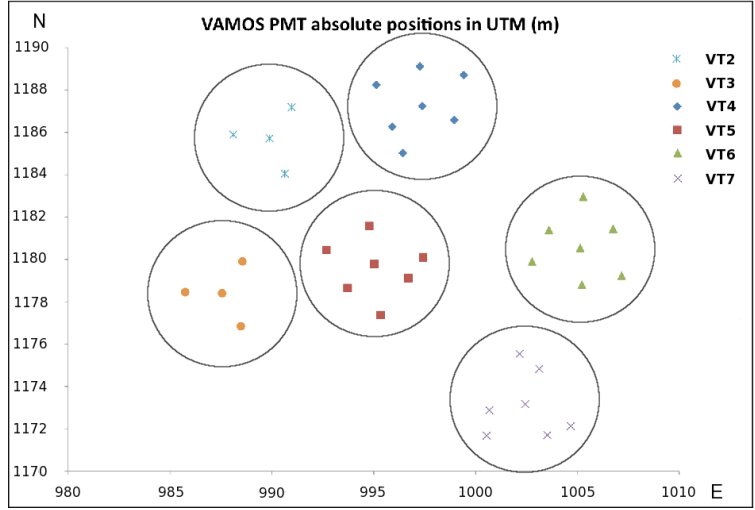}\\
 \parbox[t]{\textwidth}{
  \caption{Positions of the VAMOS entire array (respect to a reference point represented by a rock at the base of Sierra Negra mountain) based on a GPS survey using UTM coordinates
  \protect\footnotemark{}. The absolute position of the entire array based on UTM coordinates is known with an accuracy of 1~m. The different symbols used for the PMT positions correspond to different 
WCDs (see legend). X and Y axis point towards geographic East and North, respectively.}\label{PMT-position}}
\end{figure}

The VAMOS PMTs  were the 8-inch Hamamatsu 10-stages R5912SEL re-used from the Milagro experiment \citep{Milagrito}. Each PMT was equipped with the cone-shaped light collector 
already introduced for Milagro (see fig.~\ref{tank}). The six WCDs of VAMOS were built using different numbers of PMTs, in particular 4 WCDs had 7 PMTs while the other 2 WCDs contained 4 
PMTs (the planned configuration of the HAWC WCDs) as seen in fig.~\ref{PMT-position}. 
\footnotetext{The UTM system divides the Earth between $80^{\circ}$S and $84^{\circ}$N latitude into 60 zones, each $6^{\circ}$ of longitude in width, and uses a secant transverse Mercator 
projection in each zone.} 
\begin{figure}[!htb]
 \centering
  \includegraphics[width=0.6\textwidth]{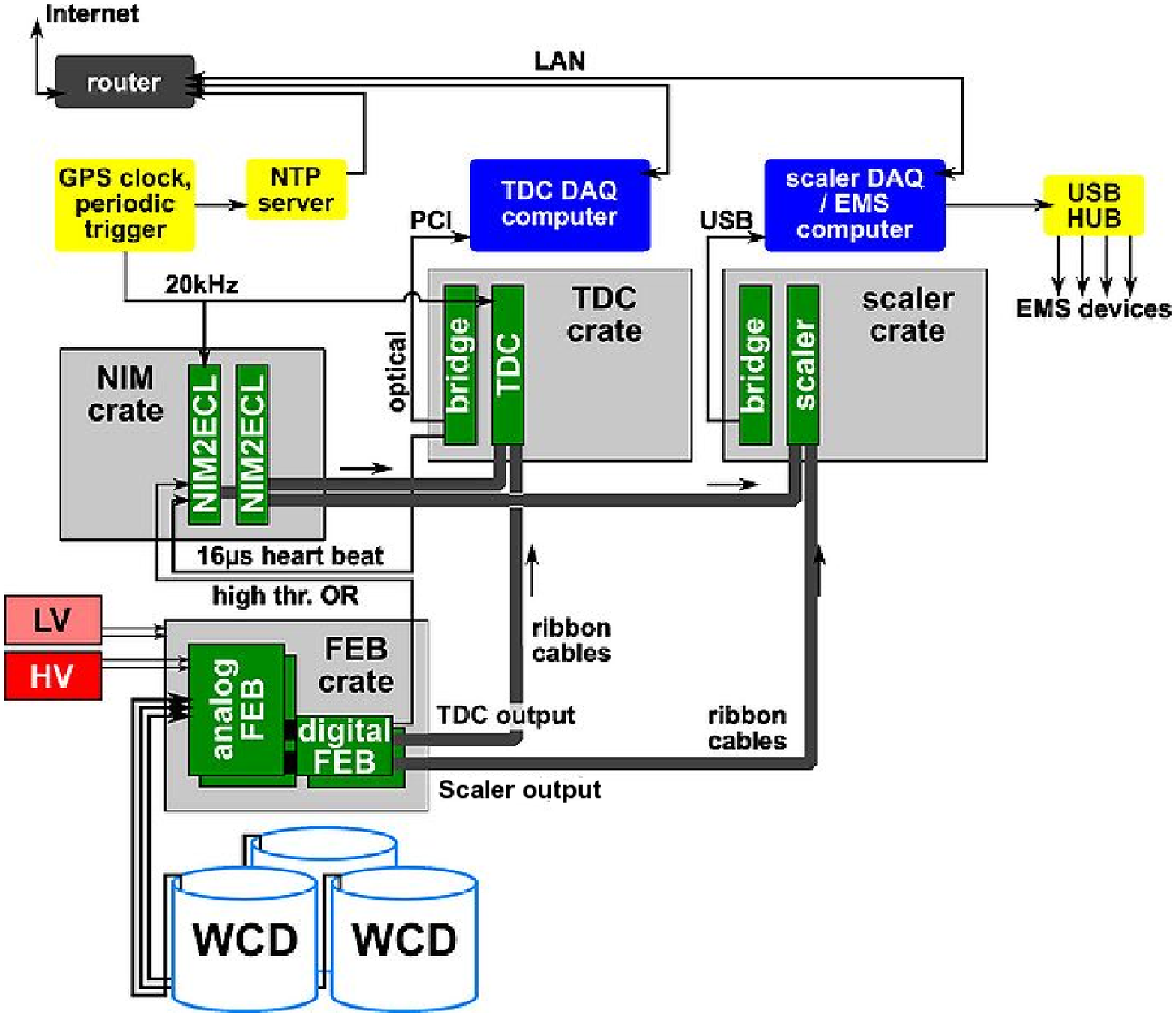}\\
  \caption{A schematic view of the two branches of VAMOS DAQ described in the text.}\label{Tdc_daq_vamos}
\end{figure}
The Cherenkov light emitted by the air shower components in the WCDs was recorded by the PMTs. The electrical signals  of VAMOS PMTs were processed by the Front End Board (FEB), 
inherited from Milagro \citep{Milagrito}. The FEB discriminators translated the analog PMT signal information into digital time stamps that encode the signal time and amplitude as Time over 
Threshold (ToT). A low threshold of 30\,mV, corresponding to 1/4 of photoelectron (PE), and a high threshold of 80\,mV, corresponding to 5\,PEs, were set in the discriminators. These 
thresholds resulted in two and four edge events encoding the ToT \citep{Milagrito}. The timestamps were digitized through a CAEN VX1190 time to digital converter (TDC) with 128 channels 
which is read out with a frequency of 20\,kHz (see fig.~\ref{Tdc_daq_vamos}). The TDCs were set up to digitize all PMT hits in a continuous mode (with no dead time due to the readout) and the 
event selection was made through the software. The data were accumulated in a PC via an optical Peripheral Component Interconnect (PCI)  to a CAEN Versa Module Europa (VME) bridge. 
The limitations on the DAQ bandwidth due to the PCI readout were overcome in the final HAWC design by using single board computers in the VME to preprocess the data and send it out via 
Gigabit Ethernet. The TDC-based DAQ system is further referred to as {\it the main DAQ} (see fig.~\ref{Tdc_daq_vamos}). The main DAQ system of VAMOS and HAWC has been designed to 
detect air shower events in the range of energy between 100\,GeV and 100\,TeV \citep{HAWCICRC}. A second branch of the data acquisition system, the scaler DAQ, was installed to increase 
the sensitivity of the detector to gamma-ray and cosmic-ray transient events (such as a Gamma-Ray Burst or a solar flare) in the GeV--TeV range using the single particle or scaler technique 
\citep{HAWCgrb}. This scaler-based DAQ system monitored the low threshold trigger of individual PMTs recording the rate of hits with more than 1/4 of PE every 10\,ms. While the scaler 
technique is not able to provide directional information about the showers, it is complementary to the event reconstruction analysis. In the following section we discuss the effect of the 
environmental variable changes on the rate of this acquisition system. 

\section{Effects of environmental variability on the PMT counting rate}
\begin{figure}[!htb]
\centering
\includegraphics[width=0.7\textwidth]{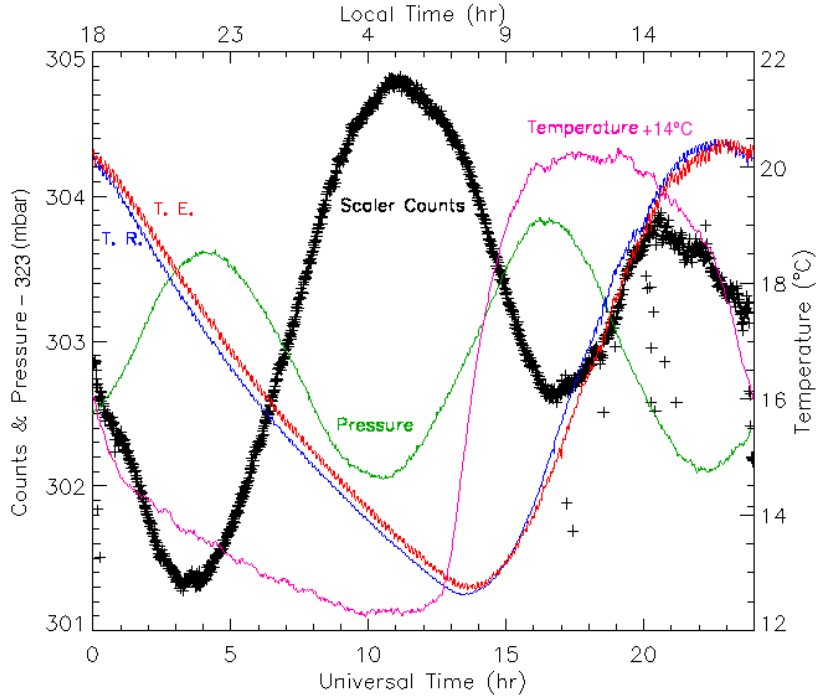}
\caption{Mean daily variation of the average PMT rate recorded by VAMOS scalers (counts/10 ms). Black crosses show the rate averaged over 30 PMTs and 41\,days with a 1\,minute 
resolution. Also shown are the ambient average pressure P (shifted by 323\,mbar, green) and average temperature T (shifted by 14$^\circ$C, pink), the internal room $T_{r}$ (blue) and 
electronics average temperatures $T_{e}$ (red). At the top of this plot we indicate the local time while at the bottom we show the corresponding universal time. In this plot we can see that the 
ambient pressure variation affects the scaler counting rate significantly more than the other environmental variables considered.}\label{total} 
\end{figure}

The components of a cosmic-ray shower and hence the PMT rates recorded by the scaler system are impacted by the environmental pressure and temperature. 
The Environmental Monitoring System\footnote{The environmental monitoring refers to the stand alone data system that records and monitors environmental parameters such as temperature, 
pressure, weather, voltages, etc.} (EMS) installed on VAMOS monitored the ambient pressure ($P$) and temperature ($T$), as well as the room temperature inside the electronics trailer 
($T_r$) and the temperature in the electronics rack ($T_e$). At Sierra Negra $P$ and $T$ have a regular behavior, with $P$ showing a 12\,hr cycle, with minima around 4:00 and 
16:00 (see fig.~\ref{total}) and maxima around 22:00 and 10:00 (local times). The mean ambient temperature showed a daily cycle with two phases: during
the night a regular decreasing trend, from $0^{\circ}$C at 20:00 to $-2^{\circ}$C at 6:00, is present; whereas during the day a ``bump'' is observed with a fast increase from $-2^{\circ}
$ to $5^{\circ}$C, an extended crest from 9:00 to 17:00 followed by a decreasing phase (fig.~\ref{total}). Also the indoor temperatures (fig.~\ref{total}) $T_r$ and $T_e$
presented a daily cycle with a slowly decreasing trend during the night reaching a minimum of  13$^{\circ}$C around 7:00 and a faster increasing trend during the day towards a well
defined peak of  $20^{\circ}$C around 16:00. Fig.~\ref{total} compares the behavior of the environmental variables introduced above with the average PMT rate. Muons are the main component 
of secondary air shower particles that impact the scaler count rate. The muon production rate is influenced by the density of the atmosphere at the relevant altitude (the 
maximum of correlation between muon rate and temperature coefficients is expected for the altitude regions of atmosphere tropopause and stratosphere \citep{Petkov}\citep{Wolfendale}), while 
the muon absorption rate correlates with the column density above the detector, which in turn is connected with the ambient pressure and temperature at the ground. Fig.~\ref{total} shows that 
the PMT rate followed a two-peak cycle, clearly in a counter-phase with the pressure. The second most significant effect, responsible for the larger rate variations during the night time, is the 
effect of the ambient temperature. Overall, the counting rate, largely affected by the muon absorption, can be well described by a parametric function of the locally measured pressure and 
temperature.

\section{Variation of scaler counting rate during a Forbush decrease event.}
\begin{figure}[!htb]
\centering
\includegraphics[width=0.9\textwidth]{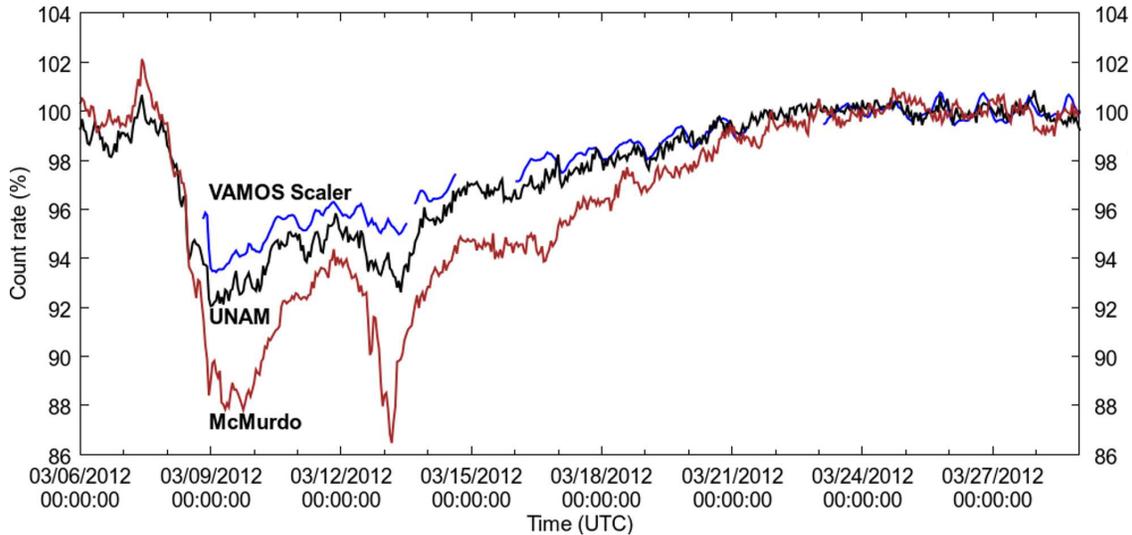}
\caption{In blue the scaler rate (corrected by pressure) from channel 10 of VT6 (see fig.~\ref{PMT-position}), in black the data from the neutron monitor of Mexico City (UNAM) and in red the data of the neutron monitor of the McMurdo observatory. The maximum of Forbush decrease is registered on March 9.}\label{fig1}
\end{figure}

A Forbush decrease (FD) event generally indicates a decrease in the cosmic-ray count rate caused by transient interplanetary events which are strictly related to a solar mass ejection activity 
\citep{hilary}. It is characterized by a sudden onset, a maximum depression reached within about one day and a more gradual recovery \citep{Lockwood}. On March 8, 2012 a major FD 
occurred as a result of the March 7 X5 solar flare (AR1429, N17E29) and associated coronal mass ejection \citep{hilary}. The FD was registered by the world-wide network of neutron monitors 
(NMs). The decrease reached its maximum early on March 9 followed by a two-week recovery. The VAMOS scaler DAQ system was not operating during the initial onset of the FD. However data 
was taken during the maximum depression and the full recovery. To observe the effects of solar modulation of galactic cosmic rays, we corrected the PMTs scaler rates considering the effect of 
pressure variability explained in section 4. The correction used is described by the following equation: 
\begin{equation}
R_{corr}=R_{meas} - R_{P}  + <R_{meas}> ,
\label{scaler_corr} 
\end{equation}
were $R_{corr}$ is the corrected scaler rate, $R_{meas}$ is the single PMT measured rate and $R_{P}=a\cdot P+b$ is the linear function used to describe the relation between the scaler rate and the 
measured atmospheric pressure (P).
The decrease in the VAMOS scaler counting rates was observed by 13 different channels. We report the 
analysis of scaler data recorded by channel n.10. From the data collected during the Forbush decrease event, we obtained the values of $a=-0.948(\pm3.2\%)$ $[hits/10\,ms][g/
cm^{2}]^{-1}$ and $b=825.335(\pm2.3\%)$ $[hits/10\,ms]$ for the selected channel. The corrected VAMOS rate measurements were compared to those of two neutron monitors: the Mexico City 
cosmic-ray observatory (at the same latitude of VAMOS prototype) and the McMurdo \citep{McMurdo} observatory (at the latitude of $78^{\circ}$ south). The Mexico City cosmic-ray observatory, 
located on the UNAM\footnote{Universidad Nacional Autonoma de M{\'e}xico} campus, is an array of cosmic-ray instruments that includes a neutron monitor (NM). Fig.~\ref{fig1} shows the corrected 
hourly count rates recorded by the channel n.10 of VAMOS scaler DAQ, the Mexico City NM and the McMurdo NM. This plot covers a period of 23\,days from maximum depression to full recovery.  A 
first analysis suggests that the features in the VAMOS scaler data replicate those in the Mexico City NM data, as well as the same features in the McMurdo NM data.  All instruments appear to have 
returned to their normal count rates by March 23, two weeks after the onset. McMurdo NM exhibited a much greater decrease compared to the VAMOS channel n.10 and Mexico City NM due to its low 
geomagnetic cutoff \citep{Smart}. The discrepancy between VAMOS and Mexico City count rate can be attributed to the different detection methods, because we are not expecting differences due to 
the geomagnetic and atmospheric cutoffs for these two detectors. Specifically, the VAMOS PMTs rate is affected by the generation and transport of the secondary muons and electromagnetic particles 
while the NM rate is affected by the production and transport of secondary neutrons. The amplitude of decrease is expected to be greater for the high latitude stations with smaller signals and 
increasing geomagnetic and atmospheric cutoffs. The exact FD spectrum dependence of atmospheric cutoff of VAMOS  is unknown. This dependence will be evaluated for HAWC, which 
will detect more FDs and ground level enhancements\footnote{A ground level enhancement is defined as a sharp increase of small duration in the counting rate of ground-based cosmic ray detectors 
caused by the accelerated charged particles from the Sun to the energies sufficiently high to be recorded at Earth} over its lifetime.
\section{VAMOS TDC data analyses}
\subsection{Event reconstruction and comparison with Monte Carlo simulation}
\begin{figure}[h!]
\centering
\begin{tabular}{cc}
\includegraphics[width=0.45\textwidth]{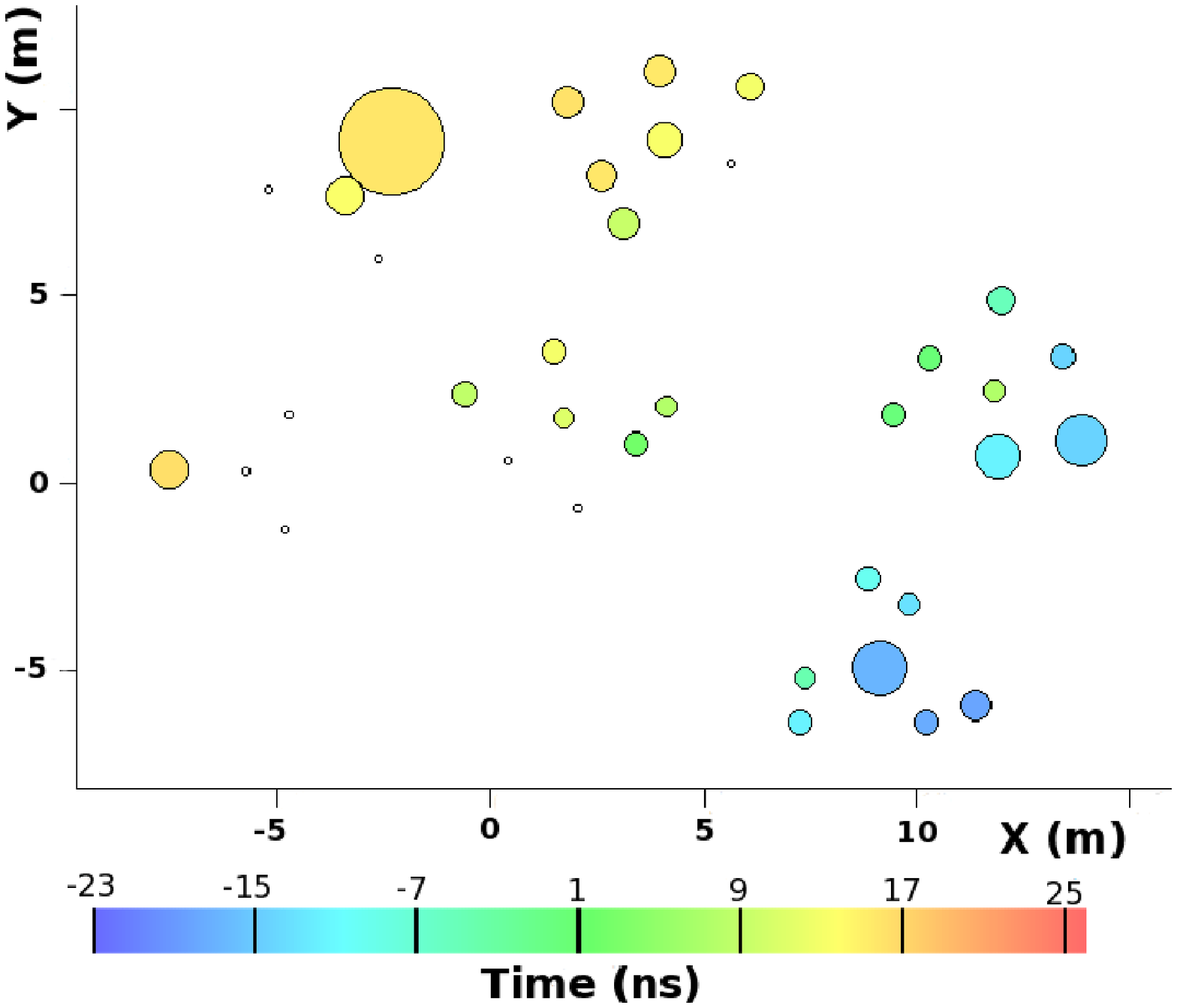}  &
\includegraphics[width=0.45\textwidth]{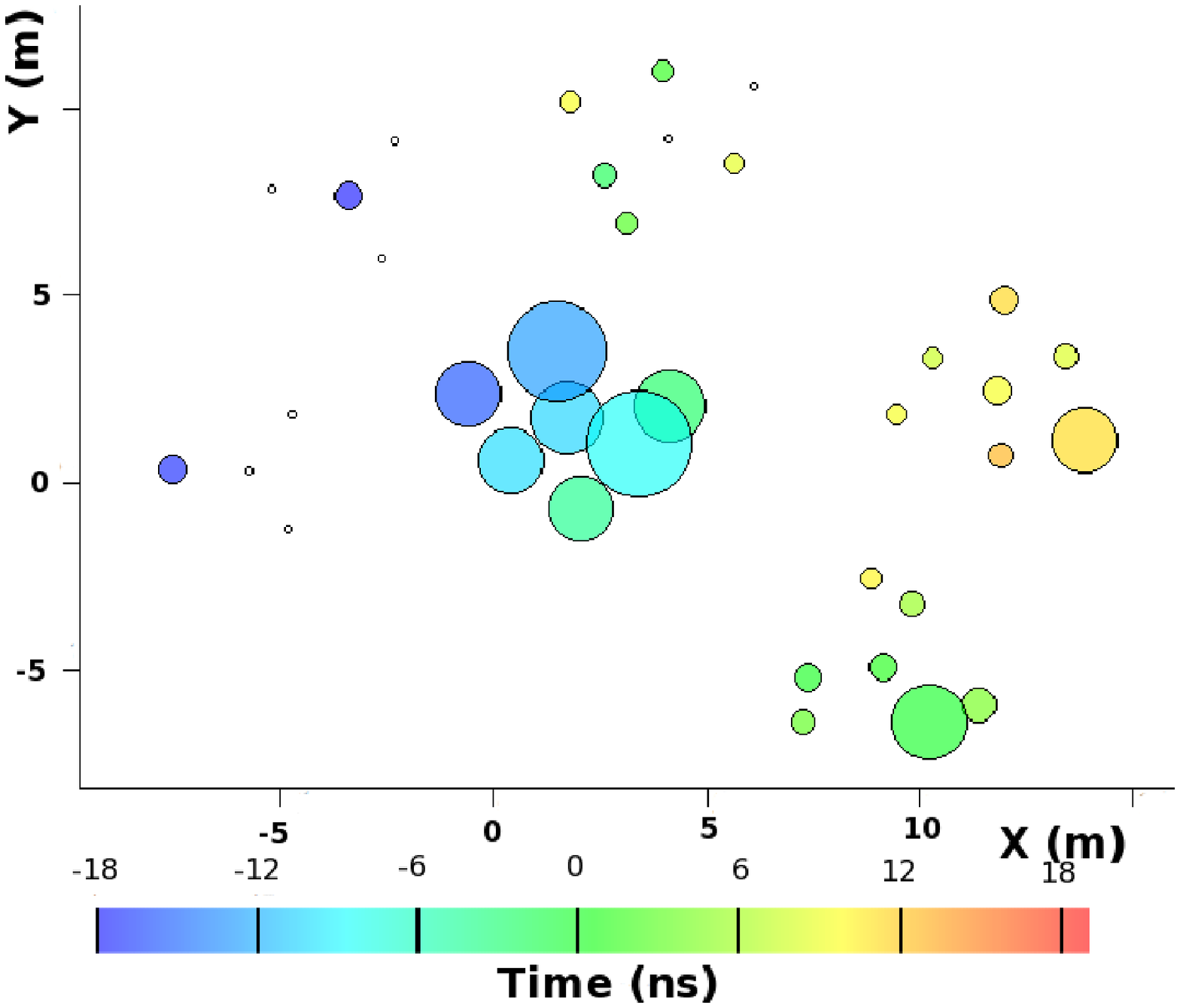}
\end{tabular}
\caption{Examples of air shower events recorded by VAMOS. Circles are drawn at the positions of the PMTs. The area of the circles is proportional to the charge observed by the PMTs. The color 
scale indicates the time of the hit relative to the trigger time. In the left plot the shower plane hits the bottom right corner first. In the right plot the shower arrives from the left side.}\label{event-display}
\end{figure}

VAMOS was operated on a regular basis starting on October 1, 2011 and ceased operations on May 24, 2012. The collected dataset includes 1349 data runs with a total live time of 83\,days (89 min per run on average). The data were processed using software to reconstruct air showers, resulting in a sample of about 3.2\,billion events. Only events with 3 or more hit 
WCDs were used in the analysis. Examples of air shower events detected by VAMOS are shown in fig.~\ref{event-display}.
\begin{figure}[!htb]
\centering
\includegraphics[width=0.8\textwidth]{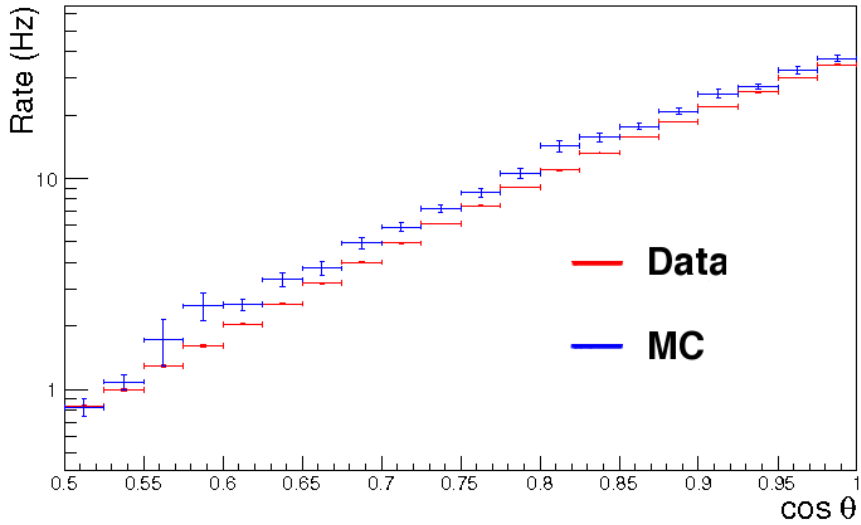}\\
\caption{Zenith angle distribution of reconstructed events recorded by VAMOS (red) in comparison with Monte Carlo predictions (blue). The data sample shown was obtained on 
March 28, 2012 using 30 PMTs (1.33\,hr of live time). Events reconstructed with less than 20 detector channels and less than 3 WCDs were excluded. The error bars show statistical errors.}
\label{MC-data-zenith}
\end{figure}
The data collected from VAMOS were used to validate the simulation models and reconstruction software developed for HAWC. The HAWC reconstruction software includes routines for finding the 
position of the shower core and direction of the shower axis \citep{Marinelli}. The core fitting is done in two stages: the center-of-mass core fitter provides a crude estimate of the core position, which is 
then improved by the Gaussian core fitter. While the first stage computes the charge-weighted center-of-mass of an air shower event, the second one is an iterative fit of the PMT charge distribution on 
the X-Y plane to a two dimensional Gaussian model. The PMT charge has been obtained using the average charge calibration curves established for the Milagro experiment. The HAWC 
experiment will have a dedicated calibration system to measure charge and timing of each PMT individually. The shower axis reconstruction is using the hit arrival times and includes two stages - 
the Gaussian plane fitter and the likelihood plane fitter. The core fitting routines and the Gaussian plane fitter are an adaptation of the corresponding Milagro reconstruction tools \citep{milagro2006}. 
The likelihood fitter was developed specifically for HAWC \citep{HAWCgrb}. The AERIE (Analysis and Event Reconstruction Integrated Environment) framework \citep{sensi} was used to integrate 
the individual reconstruction routines together. Considering the limited dimensions of the VAMOS prototype no gamma-hadron separation algorithms were applied to the reconstructed shower 
events. The reconstructed events were compared to the output of a Monte Carlo simulation of cosmic-ray air showers. Multiple species of primary cosmic-ray were simulated using CORSIKA 
\citep{corsika92}. In particular protons, He, C, O, Ne, Mg, Si and Fe have been taken into account. An $E^{-2.63}$ spectrum was considered for these species with the rates normalized to the 
measurements of the CREAM II experiment \citep{CREAMII}. The CORSIKA package computed also the interactions of primary cosmic-ray and the propagation of secondaries down to the HAWC 
altitude. The detector response, including interactions of the shower particles in the WCDs, Cherenkov light production, propagation, and detection by PMTs, were modeled using a dedicated 
package based on Geant4 \citep{geant4} (HAWCSim). A 10\,kHz random background was added to each PMT to model the effect of the PMT dark noise, radioactivity, and low energy cosmic-ray 
particles coincident with the simulated showers. The simulation chain followed the same reconstruction modules used with the experimental data. However, the data are processed with a event 
readout window of $2~\mu s$ and a trigger time window of $100~ns$ while the Monte Carlo simulations have a event noise window of $1.5~\mu s$ and a trigger time window of $190~ns$.
\begin{figure}[!htb]
\centering
\includegraphics[width=0.8\textwidth]{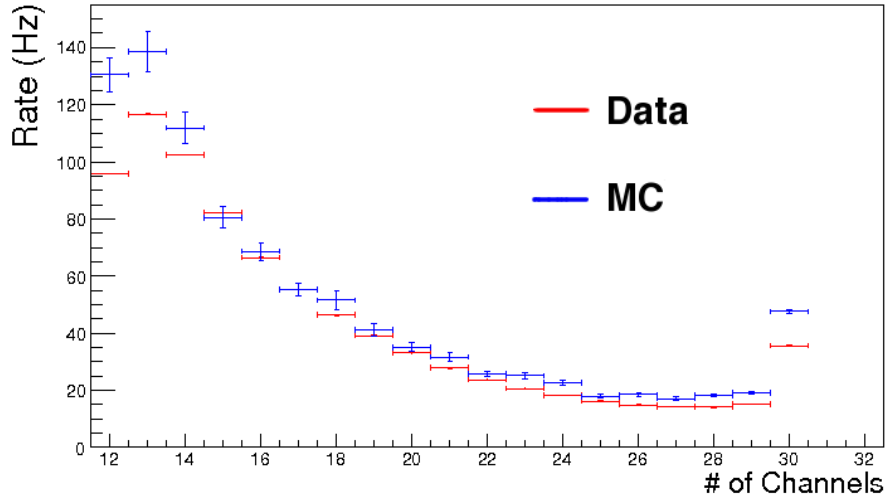}\\
\caption{Channel multiplicity distribution of reconstructed air shower events recorded by VAMOS (red) in comparison with Monte Carlo predictions (blue). Events reconstructed with 
less than 12 hits have been excluded. The error bars show statistical errors.}\label{MC-data-multiplicity}
\end{figure}
\begin{figure}[!htb]
\centering
\includegraphics[width=0.7\textwidth]{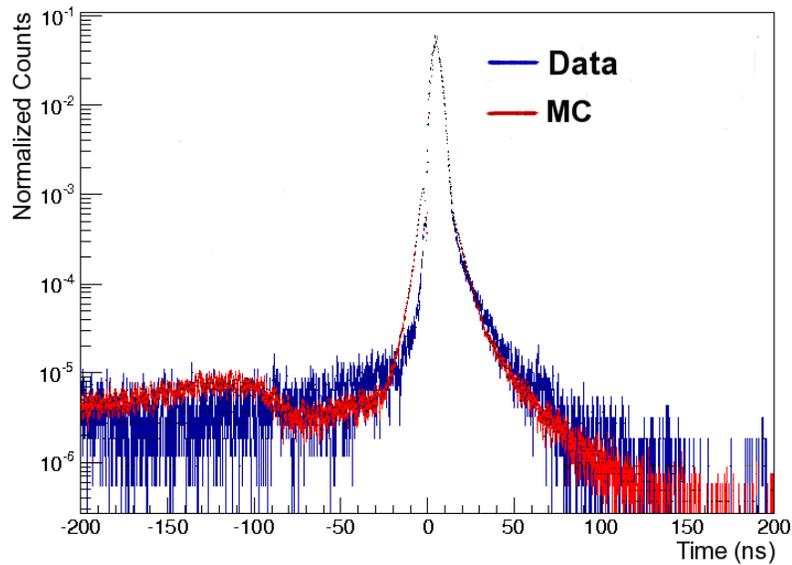}\\
\caption{The time residual distribution observed with real data (red) and simulation (blue). The peak of the time residuals is centered at 5 ns because we considered a plane shower front. Events reconstructed with less than 15 detector channels or with less than 10 PEs have been excluded.}\label{MC-data-res-time}
\end{figure}
Fig.~\ref{MC-data-zenith} shows the zenith angle distribution of the reconstructed events in the VAMOS data and compares it to the Monte Carlo simulation. The data and the MC predictions 
are in a reasonable agreement both in terms of absolute event rate and shape of the distribution within systematic uncertainties of the simulations. Fig.~\ref{MC-data-multiplicity} shows the channel 
multiplicity (number of channels hit) distribution of the reconstructed events. The data are in agreement with the simulation for intermediate values of channel multiplicity $14<N_{ch}<30$, with a 
20\% discrepancy for $N_{ch}<14$, where the selection of trigger time and readout windows may be significant, and at the saturation value $N_{ch}=30$. Overall, the observed event rate is 11\% 
below the predicted rate. The distribution of the time residuals (the difference between the hit arrival time and the time expected from the shower fit) is shown in fig.~\ref{MC-data-res-time}. The 
observed distribution closely matches the curve predicted by simulations. The good agreement in the width of the distribution indicates that the PMT time pedestals were well calibrated and the 
shower plane fit behaved as expected.

\section{Search for Gamma-Ray Bursts}

Gamma-ray bursts (GRBs) \citep{Kouveliotou} are distant transient sources of gamma rays
whose observed spectra can be affected by interactions with extragalactic background light (EBL) 
as well as propagation effects at the source.
The amount of EBL at various redshifts traces the star formation history of the Universe.
Building an accurate model of the EBL density is currently a subject of active research.
For typical GRB redshifts ($z \geq 1$), the EBL suppression should lead to a high energy spectral cutoff at around 100\,GeV
\citep{gilmore}. Fermi LAT \citep{Fermi} observations showed that GRB spectra sometimes extend beyond 30\,GeV
and into the energy range sensitive to the EBL cutoff, thus allowing one to constrain EBL models at high redshifts \citep{Abdo}.
On this matter, the highest energy measured by Fermi LAT for a photon emitted by a GRB was 95\,GeV \citep{Ackermann}.
Extending these observations to higher energies has been difficult due to limited size of the LAT.
Ground-based experiments have not so far reported a conclusive detection of a GRB,
which is due to high energy threshold (for air shower arrays) or small field of view (for IACTs).
The high altitude of the HAWC site offers new opportunities to 
observe 100\,GeV gamma rays via direct detection of gamma-ray-induced air showers.
Prospects for GRB observations with HAWC have been thoroughly discussed in \citep{HAWCgrb}.
Due to its smaller size, VAMOS only provided about 15$\%$ of the sensitivity of the full array
when we used the alternative scalers DAQ (sensitivity scales as $1/\sqrt{N_{PMT}}$)
and about 3$\%$ for the main DAQ analysis where angular resolution and gamma-hadron separation have a significant effect.
Nevertheless, primarily thanks to the high altitude, the sensitivity was comparable to that one of the Milagro experiment. 
\begin{figure}[!htb]
  \centering
  \includegraphics[width=0.7\textwidth]{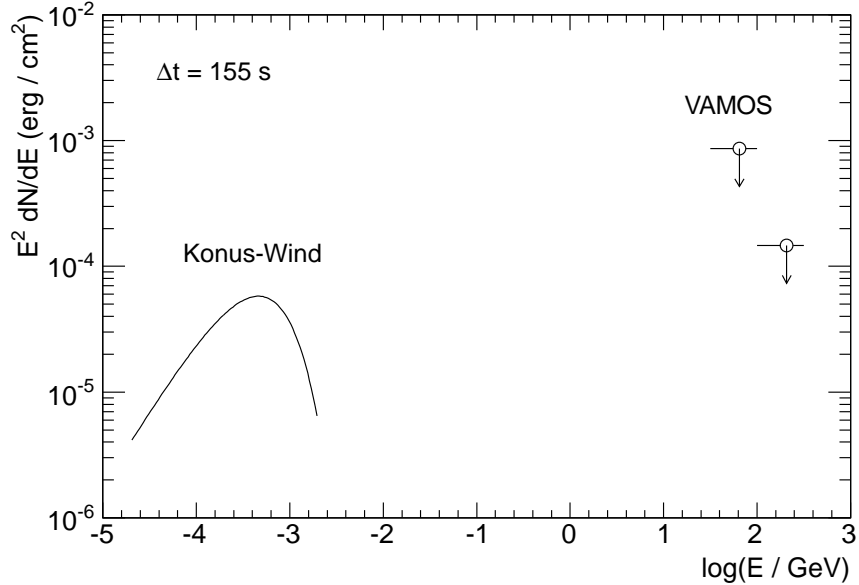}
  \caption
          {Upper limit on high energy emission from GRB 111016B imposed by VAMOS main DAQ data.
          The limit is given at 90\% confidence level for two energy bands (31.6 to 100\,GeV and 100 to 316\,GeV).
          The spectral fit reported by Konus-Wind \citep{Fredericks} is shown for comparison.}\label{grb111016b}
\end{figure}
Based on GCN circulars\footnote{http://gcn.gsfc.nasa.gov/gcn3\_archive.html}, we selected the two most intense GRBs that occurred during VAMOS operation in its field of view (at zenith angle $< 
45^{\circ}$): GRB~111016B and GRB~120328B. Both bursts were announced by IPN \citep{Palshin1} \citep{Palshin2}, with a gamma-ray fluence reported by Konus-Wind of 10$^{-4}$ erg/cm${^2}$ 
in both cases. In addition to low energy gamma rays, Fermi reported a detection of high energy emission from GRB 120328B by LAT \citep{Vianello}.
For VAMOS, GRB~111016B had a zenith angle of 32$^{\circ}$, while GRB~120328B was at a less favorable angle of 41$^{\circ}$.
No redshift information is available for these GRBs. We analyzed the VAMOS data for GRB~111016B
to establish if an intense high energy emission was present. The number of air showers detected during a 155~s time interval around the GRB
(including 5 s before T0 = 22:41:40 UT plus the reported GRB duration) and reconstructed within 6$^{\circ}$ from the GRB position
was compared to the background estimate based on the event rate in the same angular bin during a 7\,hr period including the GRB.
A negative fluctuation of $0.6\,\sigma$ was found. A 90\% C.L. upper limit on the number of signal events was derived following the method of Feldman and Cousins \citep{feldman}.                                 
The limit was then converted to flux units using a Monte Carlo simulation of the detector response.
The limits computed for two different energy bands are presented in fig.~\ref{grb111016b}.
Assuming a power law spectrum with a cutoff at 100\,GeV, the upper limit on E$^2$ dN/dE at 65\,GeV is 8.6 $\cdot$ 10$^{-4}$\,erg/cm$^2$.
For a spectrum extending up to 316\,GeV, the limit on the $>$100\,GeV emission is 1.5 $\cdot$ 10$^{-4}$\,erg/cm$^2$ at 208\,GeV.
A similar analysis was applied to VAMOS data for GRB~120328B.
Shower events were selected within a 7$^\circ$ radius bin centered at a location corresponding to the center of the
improved IPN error box (RA, Dec = 229.202$^\circ$, +24.818$^\circ$)\footnote{based on data from K. Hurley [private communication]}. 
A $+2\sigma$ fluctuation was found in a 30\,s time window following GRB onset (06:26:23 UT).
Consequently, and also due to a less favorable zenith angle, the obtained limits are weaker than for GRB~111016B:
$3.3 \cdot 10^{-3}$\,erg/cm$^2$ at 141\,GeV (100 - 200\,GeV band) and $1.4 \cdot 10^{-3}$\,erg/cm$^2$ at 283\,GeV (200 - 400\,GeV band).
The limits have been corrected by a factor of 1.6 to account for systematic uncertainties in the signal detection efficiency (following the prescription of J. Conrad et al. \citep{PhysRevD.67.012002}
and assuming a 50\% uncertainty).
These data complement the spectral measurements made at lower energies by Fermi.

\section{Summary}
VAMOS served as a pathfinder for the HAWC gamma-ray observatory, verifying and improving the design and logistics of the HAWC Water Cherenkov Detectors construction. 
This paper gives a description of VAMOS, including mechanical design, electronics and data acquisition system and summarizes the analyses performed on the VAMOS data. 
VAMOS had two independent data acquisition systems which were operated between October 2011 and May 2012 with 30\% live time. The data collected with the scaler-based DAQ system have 
been used to study the effects of environmental parameters, such as local temperature and pressure, on PMT rates. The data, corrected for the pressure changes, showed a clear signature of the 
March 2012 Forbush decrease and were considered in qualitative agreement with the observations by the Mexico City and McMurdo neutron monitors. The data collected with the main TDC-based 
DAQ system was analyzed using HAWC reconstruction software, and the key quantities and distributions were found to be consistent with Monte Carlo simulations within systematic uncertainties 
of the simulations. The data containing the bright bursts GRB~111016B and GRB~120328B were examined for signs of high energy gamma-ray signal. No statistically significant signal was found in 
the VAMOS data and upper limits on the high energy emission were presented.
\section*{Acknowledgments}
We gratefully acknowledge Scott DeLay and Federico Bareilles for their dedicated efforts 
in the construction and maintenance of the VAMOS prototype. This work has been supported
by: the National Science Foundation, the US Department of Energy Office
of High-Energy Physics, the LDRD program of Los Alamos National Laboratory,
Consejo Nacional de Ciencia y Tecnologia (grants 55155, 103520, 105033,
105666, 122331, 194116 , 132197 and 179588), Red de Fisica de Altas Energias, DGAPA-UNAM
(grants IN110212, IN105211, IN108713, IN121309, IN115409, IN111612, IN112412 and IG100414-3), 
VIEP-BUAP (grant 161-EXC-2011), Luc-Binette Foundation UNAM Postdoctoral Fellowship, 
the University of Wisconsin Alumni Research Foundation, and the Institute of Geophysics and Planetary Physics 
at Los Alamos National Lab.


\bibliographystyle{plainnat}
\addcontentsline{toc}{chapter}{Bibliography}
\begin{large}
\bibliography{biblio}
\end{large}
\mbox{}

\end{document}